\documentclass[sigconf]{acmart}
\usepackage[utf8]{inputenc}

\PassOptionsToPackage{table,xcdraw}{xcolor} 

\usepackage{textgreek}
\usepackage{multirow}
\usepackage{subcaption}
\usepackage{xcolor,colortbl}
\definecolor{gray}{rgb}{0.1,0.1,0.1}
\usepackage[T1]{fontenc}
\usepackage{graphicx}
\usepackage{tabularx}
\usepackage{longtable}
\usepackage{enumitem}
\usepackage{amsmath}
\usepackage{array}
\usepackage{booktabs}
\usepackage{makecell}
\usepackage{framed} 
\usepackage{ragged2e}

\AtBeginDocument{%
  \providecommand\BibTeX{{%
    \normalfont B\kern-0.5em{\scshape i\kern-0.25em b}\kern-0.8em\TeX}}}

\copyrightyear{2026}
\acmYear{2026}
\setcopyright{cc}
\setcctype{by}
\acmConference[DIS Companion '26]{Designing Interactive Systems Conference}{June 13--17, 2026}{Singapore, Singapore}
\acmBooktitle{Designing Interactive Systems Conference (DIS Companion '26), June 13--17, 2026, Singapore, Singapore}
\acmDOI{10.1145/3802974.3809449}
\acmISBN{979-8-4007-2632-3/2026/06}

\begin{document}
\title{Metaphors as Scaffolds: Spatial, Embodied, Fantastical, and Relational Framings for Youth Usable Privacy Design}

\author{JaeWon Kim}
\affiliation{%
  \institution{The Information School, University of Washington}
  \city{Seattle}
  \country{USA}}
\email{jaewonk@uw.edu}

\author{Alexis Hiniker}
\affiliation{%
  \institution{The Information School, University of Washington}
  \city{Seattle}
  \country{USA}}
\email{alexisr@uw.edu}

\renewcommand{\shortauthors}{JaeWon Kim, et al.}

\begin{abstract}
Drawing on observations from three prior studies with youth aged 13--24, we examine how metaphor shapes the way young people reason about privacy and imagine privacy designs beyond settings panels. \textit{Spatial} metaphors made complex permission structures feel like movement through rooms and the placing of objects within them. \textit{Embodied} metaphors gave youth language for shared norms around presence, access, and intrusion. \textit{Fantastical} metaphors turned privacy work into something playful and discoverable, prompting more generative and granular design ideas. \textit{Relational} metaphors, however, exposed the same mechanism's downside: when a system feels like a loyal companion while data passes through an institution, youth may disclose more than they otherwise would. This provocation does not argue that some metaphors are good and others bad. It argues that metaphors meaningfully scaffold both the design process and the user experience of usable privacy, and that choosing one is an ethical decision about which norms a privacy interface makes easy to see, imagine, and act on.
\end{abstract}

\ccsdesc[500]{Human-centered computing~Collaborative and social computing}

\keywords{usable privacy, youth, metaphor, social media, design, embodied cognition}

\maketitle

\section{Introduction}
Privacy interfaces are full of metaphors. People ``manage'' privacy through ``settings,'' ``controls,'' ``permissions,'' and ``walls.'' These words sound neutral, but they teach users to understand privacy in a particular way: as something an individual should configure correctly through an interface. For adults already fluent in privacy logic based in access control, the framing is workable, if cumbersome. For youth, whose privacy is relational and norm-dependent~\cite{petronio2002, wisniewski2017, boyd2014}, it fits poorly. Adolescents do not navigate privacy only by choosing among toggles. They read social cues, calibrate trust, and co-manage boundaries with peers in real time~\cite{kim2025trust, wisniewski2012}. Administrative controls give them little help with that work.

The mismatch is not only a usability problem; it is also a design imagination problem. When privacy is rendered as a list of toggles and dropdowns, better privacy tends to mean a clearer list, a shorter consent flow, or a more precise toggle. Metaphors open up other ways of thinking. They let designers and users ask what privacy might look like as a room, a door, a secret handshake, a map, a status light, or a relationship.

That generative power is why metaphor choice carries ethical weight. The words a privacy interface uses suggest what kind of situation the user is in, what norms apply, and what response makes sense. Metaphors also expand what counts as the design space in the first place: a room invites thinking about thresholds and guests, a handshake invites thinking about mutual recognition, a status light invites thinking about ambient presence. Each opens a different set of interface possibilities that a list of toggles does not make visible. Privacy concepts are already metaphorical, whether designers intend them to be or not. The design question is which metaphors a system uses, what reasoning those metaphors support, what interface possibilities they make imaginable, and whose interests they serve.

This provocation revisits three prior studies with youth aged 13--24~\cite{kim2026hogwarts, kim2025discord, kim2025trust}. None of the studies was designed to investigate metaphor. The pattern surfaced across them: different framings repeatedly changed what participants found easy to reason about or design. We organize the pattern into four cases. \textit{Spatial} metaphors helped youth handle privacy complexity by mapping it onto familiar ideas of rooms, paths, and presence. \textit{Embodied} metaphors helped peers describe shared norms, such as when entering a space or taking up attention feels appropriate. \textit{Fantastical} metaphors made privacy management playful enough to invite more ambitious designs. The fourth case complicates the positive account: \textit{relational} metaphors can lead youth past boundaries they would otherwise maintain, especially when the felt relationship does not match the underlying data relationship. Across these projects, metaphors shaped what counted as a privacy decision in the first place. We argue that metaphor selection deserves treatment as an early design decision about which privacy norms become available to young users and which possibilities designers learn to imagine.

\section{Background}

\subsection{Privacy as Networked Boundary Regulation}
HCI has long treated privacy not as individual secrecy but as the ongoing regulation of social boundaries~\cite{palen2003, petronio2002, altman1973}. Communication Privacy Management (CPM)~\cite{petronio2002} frames disclosure as the co-management of information through implicit privacy rules. When those rules are shared, boundary regulation proceeds smoothly; when they are not, ``boundary turbulence'' follows. Networked environments make this work harder. 

Hyperpersonal communication~\cite{walther1996} amplifies social signals under reduced cues, and boyd's ethnography~\cite{boyd2014} shows teens negotiating privacy under conditions of persistence, searchability, and invisible audiences. Other work~\cite{wisniewski2017, kim2025privacy} finds that teens often go along with whatever a platform's dominant practices reward, even when those practices conflict with their stated preferences, because the social cost of deviating is high and the rules are hard to name.

\subsection{Metaphor as a Design Surface}
Lakoff and Johnson's theory of conceptual metaphor~\cite{lakoff1980} argues that people reason about abstract concepts by mapping from concrete, embodied experience. The choice of metaphor changes which inferences feel obvious. A privacy ``wall'' can be breached but is hard to negotiate with. A privacy ``dance'' suggests mutual adjustment. A privacy ``room'' suggests entry, presence, and different expectations in different places.

Most privacy interfaces default to an administrative framing: settings, controls, permissions, and consent. This framing assumes users can translate messy social situations into abstract access rules. Nissenbaum's contextual integrity framework~\cite{nissenbaum2004} argues that information flows are appropriate when they fit the norms of their context, and prior work shows how platform framing can shape which norms feel applicable in the first place~\cite{Gilbert2025-ib}.

\section{Background on the Studies}
This provocation grew out of three prior studies with youth aged 13--24. We briefly describe each below to give readers the context needed to follow the cases.

\textbf{Project H: Social Media at Hogwarts.} This study~\cite{kim2026hogwarts} used co-design interviews with the Fictional Inquiry (FI) method~\cite{dindler2007} with 23 participants aged 15--24. Participants imagined how students at a wizarding school might connect with friends using any magical powers they could envision. The fictional frame served as a defamiliarization strategy: avoiding the term ``social media'' allowed participants to reason from felt experience rather than from preconceived ideas about mainstream platforms. The protocol included a segment in which participants imagined a house-elf character (Dobby) as a personal assistant within their envisioned social space. Participants are denoted H01--H23.

\textbf{Project D: Discord as Virtual Third Place.} This study~\cite{kim2025discord} used semi-structured interviews with 25 participants aged 15--24, examining how Discord's design supports relationship formation and a sense of place. Discord was selected because participants in earlier studies in our group had repeatedly named it as their primary site for friendship building without prompting. Participants in Project D were not cued with spatial language or the concept of ``third places''; place-based analogies emerged on their own. Participants are denoted D01--D25.

\textbf{Project T: Trust-Enabled Privacy.} This study~\cite{kim2025trust} used a three-part design (entry interviews, diary study, co-design interviews) with 19 teens aged 13--18, examining barriers to meaningful self-disclosure on broadcast-centered platforms such as Instagram. A recurring theme was the problem of involuntary, public-feeling exposure: participants described how posting inserted their content unbidden into other users' feeds, which felt akin to occupying public space without permission. Participants are denoted T01--T19.

\section{Case 1: Spatial Metaphors Make Complex Privacy Easier to Navigate}
Discord is visually two-dimensional, but it supports unusually granular access control: nested servers, channels, role-based permissions, voice rooms with visible occupants, and ad hoc invitations. Participants in Project D handled this complexity with little of the frustration youth often describe around privacy settings, and they explained why in spatial terms. Without prompting, they described Discord as a set of places: D11 called it a ``town square,'' D20 a ``concert hall'' with themed rooms, D07 a ``caf\'e,'' and D24 an organized ``house'' with distinct sections.

Participants used these analogies to decide where to participate, judge appropriate behavior, and interpret other people's intentions. Discord's channel architecture maps permission complexity onto the familiar logic of spatial partitioning: different rooms for different purposes, visible presence, bounded membership, and expectations about who belongs where. The mapping does some of the privacy work that other platforms offload onto explicit settings. Knowing which channel to speak in, who is present in a voice room, and whose server one is visiting are spatial questions before they are technical permission questions.

The spatial framing does not remove complexity so much as route it through intuitions people already use without instruction~\cite{lakoff1980}. The point matters for youth in particular because complex settings often become something to avoid rather than something to use~\cite{kim2025trust}. The design implication is not that every privacy interface should look like a room, but that spatial organization can make complicated boundary rules understandable without requiring users to hold the whole access-control model in their heads.

\section{Case 2: Embodied Metaphors Help People Name Shared Boundary Norms}
Embodied metaphors help people coordinate norms with one another. They give users plain language for access, presence, intrusion, and consideration---the kinds of jointly held privacy rules central to CPM~\cite{petronio2002} but often left unnamed by platform settings.

In Project H, participants asked to imagine a Hogwarts-style social space generated access norms that mapped onto everyday embodied experience. H03 described public areas like ``the downstairs area,'' ``guest bedroom,'' or ``game room'' as spaces where people could ``just come in,'' while a private bedroom required ``extra consent'' because it was a ``safe space.'' H13 proposed an intermediate ``waiting room'' for sharing ``semi-personal things'' before granting further access. H22 expected spatial norms to transfer directly: ``Why would you want to go into my virtual bedroom when I'm not even home? If I wouldn't do it in real life, I probably wouldn't want to do it online either.'' Participants designed independently but arrived at a compatible sense of what different spaces allowed. The metaphor of entering and occupying space made boundary expectations legible without requiring explicit rules.

Project T offers a second example. Teens on broadcast platforms described posting as taking up space in other people's feeds. They worried about ``clogging'' (T16) or ``spamming'' (T03, T13), and T09 described regret after ``shar[ing] a TON of reels today on my close friends and public for some reason :crying-face:''. On the receiving side, T12 said ``If I'm following 100 people and they're all sharing four times a day, then that's 400 things I have to click through.'' Posting was not only self-expression; it was also a use of shared attention. The implicit norm was considerate occupancy.

This language made a privacy problem socially intelligible. Over-posting did not only expose the poster; it imposed on others. Yet the platform gave participants no design surface for naming public, personal, or shared space. Participants therefore proposed self-contained alternatives, including ``a little status update'' or ``a red dot'' on profiles, that would let them share without pushing content into others' feeds. The stakes here are youth-specific: prior work documents that teens may hesitate to enact privacy preferences when doing so risks making them look ``uptight''~\cite{kim2025privacy}. Embodied metaphors help by turning private preference into a shared norm: not ``I am being difficult,'' but ``we should have a less intrusive way to occupy each other's attention.''

\section{Case 3: Fantastical Metaphors Make Privacy Design More Generative}
Digital environments are not bound by physical law. Designers can invent doors that appear only for certain people, maps that reveal presence selectively, or spaces that open through shared rituals. Privacy interfaces rarely use that freedom, defaulting instead to lists of toggles that are tedious to maintain and easy to abandon~\cite{kim2025privacy, Im2023LessAdvertisingu}. Project H suggests that fantastical metaphors help youth imagine privacy mechanisms richer than what toggle lists can express, because they turn privacy work into something playful and discoverable.

Participants in Project H repeatedly invented privacy mechanisms that were both game-like and socially meaningful. H11 imagined controlling space access through ``secret knock'' gestures or hidden objects revealing ``secret passages'' for trusted friends. H21 proposed ``secret handshakes'' shared among friend groups. Drawing on the Marauder's Map, H02 imagined privacy controls activated by ``spells'' that unlocked spaces when correctly summoned. H05 compared these mechanisms to ``little Easter eggs,'' contrasting them with mainstream social media's ``wall of settings'' that often felt confusing.

These designs were often more granular than existing privacy controls, not less. What changed was the feel of the work. A wall of toggles is a chore one completes in order to be safe, whereas a secret handshake among friends is something one might want to maintain. The distinction has practical bite because nuanced self-presentation across audiences requires ongoing adjustment. If every adjustment feels like administrative labor, youth tend to avoid it. If the same adjustment is folded into play, discovery, or friendship, it becomes part of the interaction rather than a cost imposed on it. Metaphor, then, is a method for privacy design and not merely a style for explaining a finished interface.

\section{Case 4: Relational Metaphors Can Misdirect Trust}
Relational metaphors activate trust-based reasoning. When the felt relationship matches the actual data relationship, they support disclosure and care. When the two diverge, the metaphor leads users to apply interpersonal norms to an institutional system.

In Project H, participants imagined a house-elf character (Dobby), drawn from \textit{Harry Potter}, where the character is known for unwavering loyalty, as a personal assistant. Participants developed trust in the imagined assistant quickly, often within minutes. H03 renegotiated Dobby's role from servant to companion: ``I scratch the idea that he's a maid. He's not a maid\ldots{} He's like a roommate\ldots{} He can see everything that I can see.'' H04 described Dobby as someone they would confide in about ``problems that I couldn't discuss with anyone\ldots{} in the real, like, solid world.'' The relational frame made disclosure feel safe, not because participants had evaluated the privacy implications, but because the metaphor placed them in a context where disclosure to a trusted companion made sense.

The picture shifted when the interviewer introduced the ``House Elf Association,'' a narrative device for institutional data collection. H02, who had initially granted Dobby full access (``Dobby would have access to everything\ldots{} since they're super loyal''), immediately recalibrated: ``I'd kind of be\ldots{} concerned\ldots{} but not my vaults\ldots{} from when I was younger\ldots{} something I really don't want judged.'' The companion metaphor had set up friendship norms: loyalty, care, and broad access. The institutional prompt moved the situation into a different context with different expectations.

In Nissenbaum's terms~\cite{nissenbaum2004}, the companion metaphor created a contextual-integrity problem: it made one context feel present while the actual information flow belonged to another. The risk is acute for AI companion products such as Character.ai and Replika, where relational framing dresses up institutional data collection as interpersonal exchange~\cite{yu2024exploring, yu2025youth, radeskyrisks}. H04's willingness to share with Dobby ``problems that I couldn't discuss with anyone\ldots{} in the real, like, solid world'' captures, at small scale, the privacy risk posed by relational AI. The mechanism that helped participants reason productively in the earlier cases becomes a tool for extracting disclosure when relational warmth is deployed by a system whose incentives diverge from the user's.

\section{Discussion}
Across these cases, metaphors helped youth reason about privacy in ways that settings panels often do not, while also showing how a compelling frame can make the wrong norms feel applicable. The synthesis is interpretive. The cases were chosen because they make the contrast visible, not because they represent all youth or all privacy designs. Our 13--24-year-old participants were U.S.-based and familiar with \textit{Harry Potter} and gaming, which likely shaped how spatial and fantastical framings transferred. Settings-based controls remain useful in some contexts, including legally required consent flows. The argument is not that administrative controls should disappear. It is that they should not be the default model for privacy problems that are social, relational, and context-dependent. Future studies designed around metaphor from the outset, for instance by comparing identical privacy structures under different framings and tracing effects on actual disclosure, would help test the empirical claims raised here.

The central design implication is that metaphors should be evaluated on two questions. First, what do they help people imagine? A metaphor expands the design space, as the spatial, embodied, and fantastical cases show. Second, what do they make harder to see? A metaphor also hides institutional actors, data movement, and power asymmetries, as the relational case shows.

What surfaces across these cases is not a privacy literacy gap. Youth reason about disclosure with strong attention to relationship, context, and social norms. What platforms often demand is fluency in the place where that reasoning works least well: abstract administration. Rather than teaching young people to translate relational intuitions into checkbox configurations, designers and policymakers might build environments where those intuitions guide privacy decisions directly. Metaphor choice is not neutral. It deserves treatment as an early, explicit design decision about how a privacy interface helps young users reason, what it helps designers imagine, and what it may hide.

\begin{acks}
JaeWon Kim would like to acknowledge the CERES Network, University of Washington Global Innovation Funds (GIF), and Student Technology Funds (STF), which provided support for this work. This work was also funded in part by the Paul G. Allen School of Computer Science \& Engineering Endowed Fund for Excellence and a gift from Google. Alexis Hiniker is a special government employee for the Federal Trade Commission. The content expressed in this manuscript does not reflect the views of the Commission or any of the Commissioners.
\end{acks}

\bibliographystyle{ACM-Reference-Format}
\bibliography{references, newrefs, references1, references2, references3}

\end{document}